\documentclass {elsart}

\usepackage{graphicx}
\usepackage{amsfonts}
\usepackage{amssymb}

\begin{document}
\begin{frontmatter}
\title{An Improved Shashlyk Calorimeter}
\author[INR,Yale]{G.S.~Atoian},
\author[IHEP]{G.I.~Britvich},
\author[IHEP]{S.K.~Chernichenko},
\author[Yale]{S.~Dhawan},
\author[INR,Yale]{V.V.~Issakov},
\author[INR]{O.V.~Karavichev},
\author[INR]{T.L.~Karavicheva},
\author[INR]{V.N.~Marin},
\author[INR,Yale]{A.A.~Poblaguev\corauthref{email}},
\author[IHEP]{I.V.~Shein},
\author[IHEP]{A.P.~ Soldatov}, and
\author[Yale]{M.E.~Zeller}
\address[INR]{Institute for Nuclear Research of Russian Academy of
Sciences, Moscow 117312, Russia}
\address[IHEP]{Institute for High Energy Physics, Protvino 142284, Russia}
\address[Yale]{Physics Department, Yale University, New Haven, CT 06511, USA}
\corauth[email]{Corresponding author. 
{\em E-mail address:} poblaguev@bnl.gov (A.A.~Poblaguev)}
\date{27 September 2007}

\begin{abstract}
Shashlyk electromagnetic calorimeter modules with an energy resolution of about
$3\%/\sqrt{E\ \mathrm{(GeV)}}$ for $50\mathrm{-}1000\ \mathrm{MeV}$ photons
has been developed, and a prototype tested.
Details of these improved modules, including mechanical 
construction, selection of wave shifting fibers and photo-detectors, 
and development of a new scintillator with improved optical and mechanical properties are described.
How the modules will perform in a large calorimeter was determined from
prototype measurements. 
The experimentally determined characteristics 
of the calorimeter prototype show energy resolution of
$\sigma_E/E=(1.96\pm0.1)\%\oplus(2.74\pm0.05)\%/\sqrt{E},$
time resolution of 
$\sigma_T = (72\pm4)/\sqrt{E}\oplus(14\pm2)/E\ \mathrm{(ps)}$, where photon energy
$E$ is given in $\mathrm{GeV}$ units and $\oplus$ means a quadratic
summation. A punch-through inefficiency of photon detection was measured to be 
$\epsilon \approx 5\times 10^{-5}\;\;(\Theta _{\mathrm{beam}} >5\;\mathrm{mrad})$.

\end{abstract}
\begin{keyword}
Shashlyk calorimeter \sep scintillator \sep WLS fiber \sep APD \sep WFD \sep Monte-Carlo simulation
\PACS 29.40.Vj \sep 07.05.Tp
\end{keyword}

\end{frontmatter}

\section{Introduction.}

The Shashlyk technique for electromagnetic colorimetry
\cite{NIM_A320_144} has been in use for 
several years. In designing a large calorimeter for
the KOPIO experiment \cite{KOPIO} we have developed an improved Shashlyk module 
for such a device.  This paper describes the design and construction of the
module, as well as the unit's performance in prototype tests.   It is further 
development of the work described in Ref \cite{NIM_A531_467}. 

The requirements of the KOPIO experiment led to the following specifications:

\begin{itemize}
\item Energy range: $50\mathrm{-}1000$ $\mathrm{MeV}$.
\item Energy resolution: $\approx 3.0\%/\sqrt{E(\mathrm{GeV})}$.
\item Time resolution:   $\approx 100\ \mathrm{ps}\,/\sqrt{E\
  \mathrm{(GeV)}}$.
\item Effective noise per channel: $\leq 0.5\ \mathrm{MeV}$
\item Photon detection inefficiency: $\le10^{-4}$. 
\item Granularity: $\sim10\ \mathrm{cm}$.
\item Size of the Calorimeter: $4.4\times4.4$ m$^2$.
\item Operation in a magnetic field of up to 500 Gauss.
\end{itemize}

The Shashlyk approach, in which modules are constructed of lead-scintillator 
sandwiches read out by means of 
Wavelength-Shifting (WLS) fibers passing through holes in the scintillator 
and lead is appropriate for such specifications.  The 
performance level can be achieved economically with well understood and
reliable techniques.  We describe a module with significantly
improved performance over previous manifestations for which the technique is well 
proved by us, {\it e.g.} experiment E865 at Brookhaven
~\cite{NIM_A320_144,NIM_A479_349}, 
and has been adopted by others, {\it e.g.} the PHENIX RHIC detector 
\cite{NIM_A499_521}, the HERA-B detector at DESY~\cite{NIM_A461_332}, 
and the LHCb detector at CERN~\cite{NIM_A462_233}.

\section{Approach to improving the Shashlyk module}

At the outset of improving the module a detailed simulation model was
developed \cite{NIM_A531_467}. This model is based on a GEANT3
\cite{GEANT3} description of electromagnetic shower and a special
optical simulation of the light collection in the scintillator
plates. Input for the model was experimental and test beam data. The 
model output correlated well with data. Analyzing the performance of
the prototype Shashlyk module with energy resolution of about
$4\%/\sqrt{E(\mathrm{GeV})}$ %
\footnote{This module was described in details in
  Ref. \cite{NIM_A531_467}. In this paper we will refer to 
  it as to ``earlier'' Shashlyk module.}
we found, that energy resolution of about $3.0\%/\sqrt{E(\mathrm{GeV})}$
could be reached after some optimization of the module's mechanical and optical
properties (see details in Ref.~\cite{NIM_A531_467}).

We revisited the mechanical and optical construction as described below, and 
optimized the 
selection of WLS fibers and the photo-detector.
A new scintillator with improved optical and mechanical properties was
specially developed at the IHEP scintillator
facility (Protvino, Russia)~\cite{IHEP}. The corresponding improvements
to the module design were implemented in the new KOPIO Calorimeter prototype
modules, which were equipped with an Avalanche Photo Diode (APD) and Wave-Form 
Digitizer (WFD) readout.

\subsection{Mechanical construction of the improved Shashlyk module.}

\begin{figure}
\begin{center}
\includegraphics*[width=0.75\textwidth]{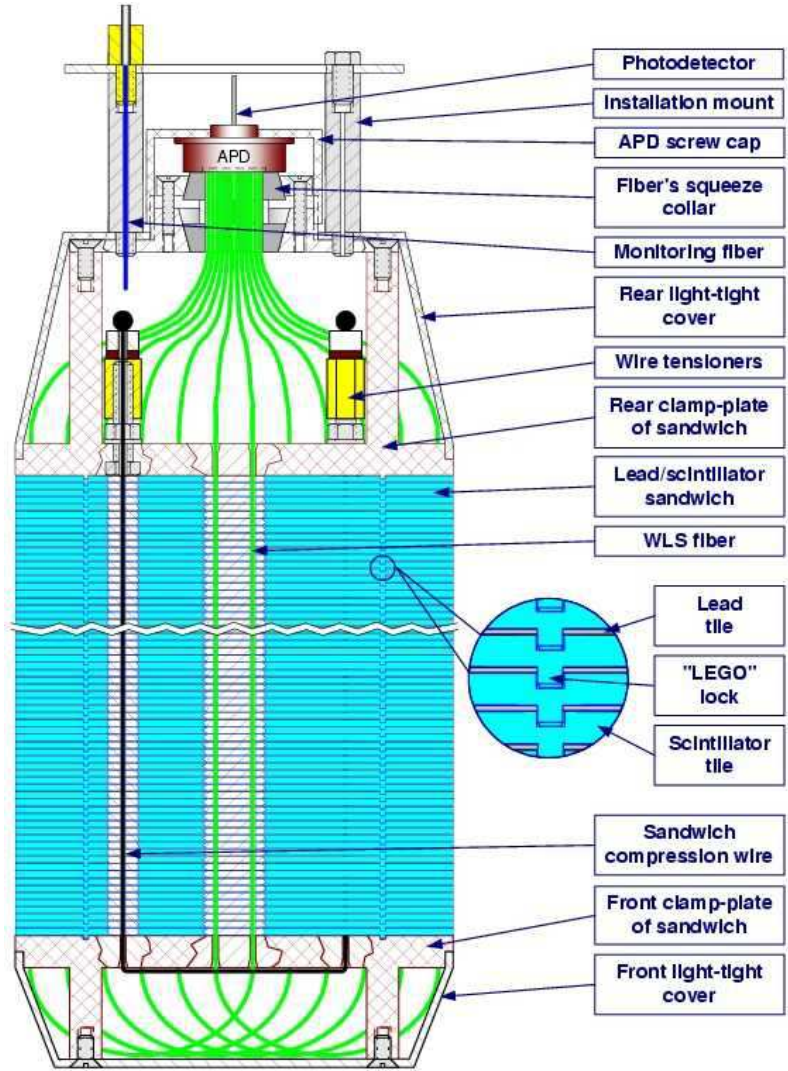}
\caption{The Shashlyk module design.}
\label{module_design}
\end{center}
\end{figure}

\begin{figure}
\begin{center}
\includegraphics*[width=0.65\textwidth]{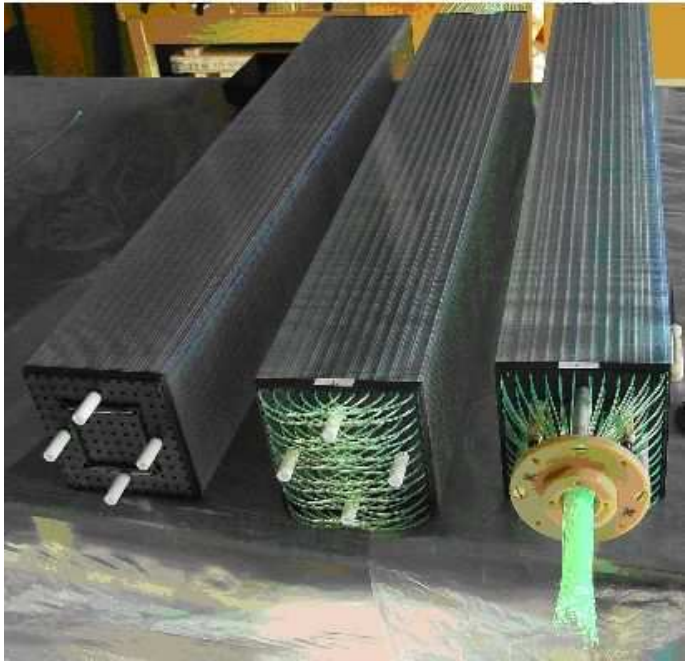}
\caption {The Shashlyk modules at different stages of assembly.}
\label{module_assembly}
\end{center}
\end{figure}

The design of the new prototype module is shown in Figures ~\ref{module_design}, ~\ref{module_assembly}.
The module is a sandwich of alternating
perforated stamped lead and injection-molded polystyrene-based 
scintillator plates. The cross sectional size of a module is 
$110\times110$~mm$^2$. There are 300 layers, 
each layer consisting of a 0.275-mm lead plate and 1.5-mm scintillator 
plate. The lateral size of a lead or scintillator plate is 
$109.7\times109.7$~mm$^2$. Each plate has 144 holes equidistantly 
arranged in a $12\times 12$ matrix, in which the spacing between the holes 
is 9.3~mm. The diameter of the holes is 1.3~mm, both in the lead 
and the scintillator plates. Inserted into the holes are 72 WLS 
fibers, with each fiber looped at the front of the module so that 
both ends of a fiber are viewed by a photo-detector. Such a loop 
(radius $\sim$2.5 cm) may be approximated by a mirror with 
a reflection coefficient of about 95\%~\cite{KURARAY}. The fiber ends 
are collected in one bunch, squeezed, cut and polished, and connected 
to a photo-detector though a small air gap. The complete stack of all 
plates is held in compression by four 1-mm stainless steel wires. 
The module is wrapped with 150-$\mu$m TYVEK paper which has light 
reflection efficiency of about 80\%.

The mechanical parameters of the module are summarized in Table~\ref{module}.

\begin{table}
\caption{Parameters of the improved Shashlyk module.}
\label{module}
\begin{tabular}{ll}
\hline
Transverse size                     & $110\times110~\mathrm{mm}^2$ \\
Scintillator thickness              & $1.5~\mathrm{mm}$            \\
Spacing between scintillator tiles  & $0.350~\mathrm{mm}$          \\
Lead absorber thickness             & $0.275~\mathrm{mm}$          \\
Number of the layers                & 300                          \\
WLS fibers per module               & $72\times1.5~\mathrm{m} = 
                                                   108~\mathrm{m}$ \\
Fiber spacing                       & $9.3\ \mathrm{mm}$           \\
Hole diameter (lead/scint.)         & 1.3 mm                       \\
Diameter of WLS fiber (Y11-200MS)   & 1.0 mm                       \\
Fiber bunch diameter                & 14.0 mm                      \\
External wrapping (TYVEK paper)     & 150 $\mu$m                   \\
Effective radiation length, $X_0$   & 34.9 mm                      \\
Effective Moliere radius, $R_M$     & 59.8 mm                      \\
Effective density                   & 2.75 g/cm$^3$                \\
Active depth                        & 555 mm~~(15.9 $X_0$)         \\
Total depth (without photo-detector) & 650 mm                       \\
Total weight                        & 21 kg                        \\
\hline
\end{tabular}
\end{table}

The module is assembled in a special assembling berth that allows one
to subject the assembled modules to cyclic longitudinal loadings up to 800 kg.
This procedure prevents subsequent longitudinal
shrinkage of the assembled modules and
provides long-term stability for the length of the module
to an accuracy of $\simeq  1\ \mathrm{mm}$.

\subsection{Improvements of the Module Geometry}

The mechanical construction of the module was revisited to minimize 
the insensitive area, to increase the effective radiation density,
and to improve the sampling ratio and transverse light collection
uniformity.
 
An important innovation in the mechanical design of the module is the
``LEGO''-type locks for the scintillator tiles shown in Fig.\ 
\ref{module_design}. These locks, four per tile, maintain the position of the 
scintillators and the 350-$\mu$m gaps, providing sufficient room for 
the 275-$\mu$m lead tiles without optical contacts between lead and
scintillator.
This mechanical structure enables removal of the 600 paper
tiles that were in earlier modules, and allows reduction of the diameter of 
the fiber holes to 1.3~mm and removal of the compressing steel
tapes at the sides of the module. Compared to the earlier 
Shashlyk module, the holes/cracks and other insensitive areas were 
reduced from 2.5\% to 1.6\%, and the module's mechanical properties 
such as dimensional tolerances and stiffness were 
significantly improved. By removing the paper tiles, the effective 
radiation length could be reduced from 4.0~cm to 3.5~cm.

\begin{figure}
\begin{center}
\includegraphics*[width=0.65\textwidth]{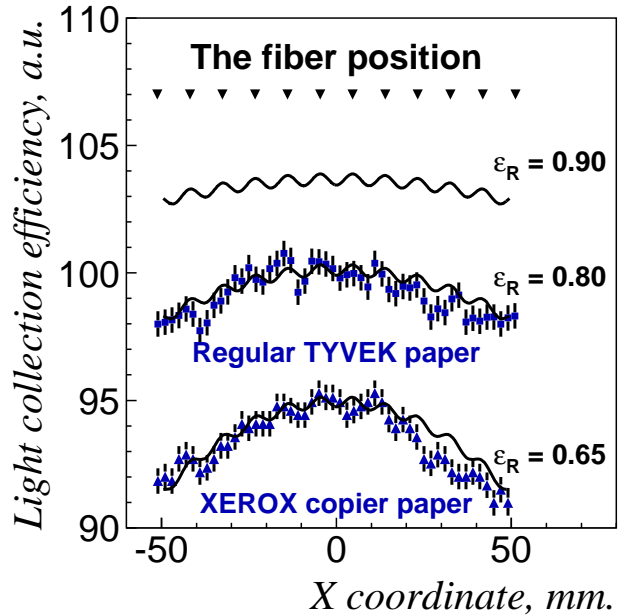}
\caption {The dependence of the light collection efficiency in the
         scintillator tile on the 
         $x$-coordinate of the point-like light source. Solid lines are 
         the simulations for the specified reflection efficiencies 
         $\varepsilon_R$ of the wrapping material.}
\label{tile_uniform}
\end{center}
\end{figure}

The sampling, i.e., the relation between thicknesses of lead and
scintillator tiles, dominates in the energy resolution of the Shashlyk module.
However, one has only limited ability to improve the ``pure" 
sampling contribution to the energy resolution of the Shashlyk 
module. Decreasing the thickness of the lead will increase the
length of the module, while the proportional decreasing  both the 
lead and scintillator thicknesses will reduce the light collection
efficiency. Nevertheless, by removing the paper between the lead 
and scintillator tiles, both the sampling could be improved and the 
length of the module could be shortened. Compared to the design of the 
previous module~\cite{NIM_A531_467}, the sampling ratio was improved 
by a factor of 1.25.

\begin{figure}
\begin{center}
\includegraphics*[width=0.65\textwidth]{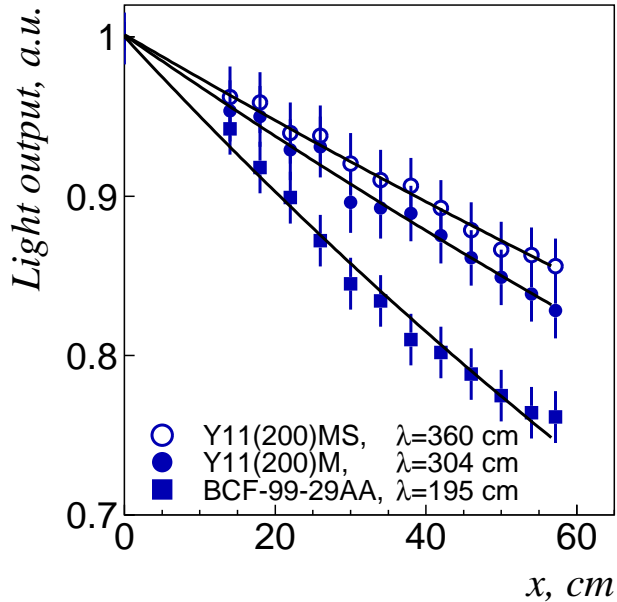}
\caption{
  The effective attenuation of the light in the fibers of Shashlyk 
  module. Experimental data (marks) are fit by the exponential dependence 
  $\exp{(-x/\lambda )}$ (solid lines), where $x$ is the distance 
  to the photo-detector and 
  $\lambda $ is the effective attenuation length.
}
\label{wls_attenuation}
\end{center}        
\end{figure}

The dominant source of non-uniformity of the light collection is
the absorption of light at the edges of a scintillator tile.  The reflection efficiency on the edges of the
scintillator tile is crucial. 
In the new module, the WLS fiber
density was effectively increased by reducing the size 
of the tiles to 10.97 cm. This allows the outer fibers to be closer to the edge of the scintillator tile than  than they were with the the ``uniform'' size of 11.16 cm for 12 fibers with 0.93-cm spacing.
In addition, the  module was wrapped with TYVEK
paper (reflection efficiency about 80\%). 
As a result, the light collection efficiency at the 
edges of the scintillator tile is only 2.3\% smaller than in the 
center of the tile for the point-like light source. In the case of a 
250-MeV photon shower, the difference is only 0.5\%, which is 
negligible compared to the energy resolution of about 6\% for such 
photons.

The experimental results for the light collection uniformity for TYVEK 
and Xerox copier papers are presented in Fig.~\ref{tile_uniform}. 
The measurements were made with a scintillator tile exposed to 
collimated $^{90}$Sr electrons.  For comparison, the simulated light 
collection efficiencies are shown. One can see that there is consistency
between the optical simulations and measurements. Further improvement 
could be achieved if Millipore paper with reflection efficiency of 90\% 
were used~\cite{refl_eff}.

\subsection{WLS Fibers}

A main concern about the WLS fibers for the Shashlyk 
readout is the light attenuation length in a fiber.  Longitudinal fluctuations of electromagnetic showers are about 3-4~cm 
(one radiation length).  The effective attenuation length in fibers, including the effect of the fiber loop
  and the contribution of the short-distance component of light, must 
be greater than 3--4 m to have this contribution to the energy resolution
be much smaller than the sampling contribution.

We have measured (see Fig. \ref{wls_attenuation}) the light attenuation in few
multi-clad WLS fibers using a  $1\times1\ \mathrm{cm}^2$ wide
muon beam penetrating the module transversely.  

  The effective attenuation 
length of 3.6~m in KURARAY Y11(200)MS fiber satisfies our requirements.
In comparison with other fibers, this commercial fiber also provides 
the best reemission efficiency of blue scintillation light, and 
has excellent mechanical properties, high tensile and bending 
strength, and high uniformity in cross-sectional dimensions. 
For example, its light reemission efficiency is a factor of 1.5 
larger than that for any Bicron WLS fiber, and the diameter for 
any round fiber is more uniform than others in that it varies by no more than $2.0\%$. 

\subsection{Scintillator}

An important contribution for the improvement of the photo-statistics 
over earlier designs of Shashlyk modules is the use of new scintillator 
tiles with an increased light collection efficiency. An optimization of 
the light yield of the scintillator tiles for the KOPIO Shashlyk modules 
has been developed and carried out at the IHEP scintillator facility 
(Protvino, Russia)~\cite{IHEP}. In the previous Shashlyk calorimeters, 
scintillator based on PSM115 polystyrene was used. The new modules 
employ BASF143E-based scintillator.

\begin{figure}
\begin{center}
\includegraphics*[width=0.65\textwidth]{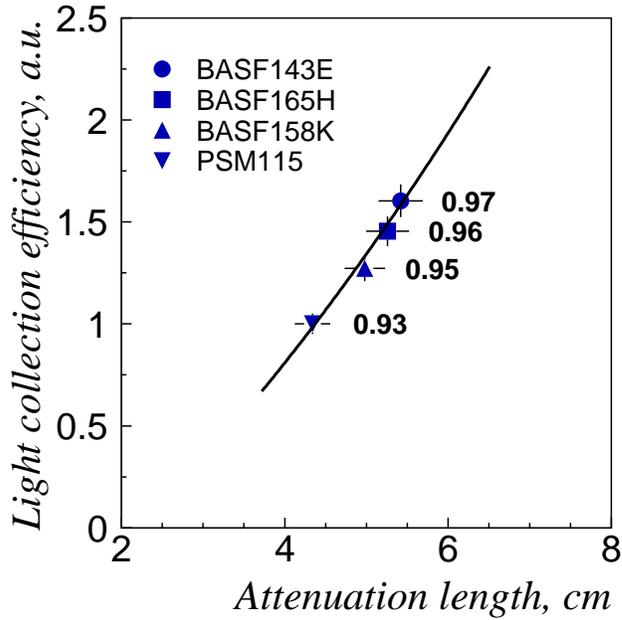}
\caption{The parametric dependence of the light collection efficiency
         on the effective attenuation length due to the reflection
         efficiency. The solid line is the result of a calculation based 
         on the model of Ref.~\cite{NIM_A531_467}.
         For each scintillator, the calculated reflection efficiency is 
         displayed.}
\label{tile_reflection}
\end{center}        
\end{figure}

Though there is no actual increase in the amount of light produced 
by a charge particle, the light collection efficiency in the new scintillator 
tile has a gain by a factor of 1.6. Because the index of 
refraction for the polystyrene-based scintillator is 1.59, only light
from total internal reflection on the large side 
of the scintillator tile can be captured by a WLS fiber.  The total 
internal reflection efficiency can potentially be as large as 100\%.  
However, this value is not reachable for realistic surfaces. Many 
reflections usually occur before light is captured and re-emitted by 
a WLS fiber. Both the effective attenuation length and the light 
collection efficiency in a scintillator tile depends on the light 
reflection efficiency. As shown in Fig.~\ref{tile_reflection}, 
the parametric dependence of the light collection efficiency on the 
attenuation length, calculated from our model of module optics 
\cite{NIM_A531_467}, is in a good agreement with the 
experimental results. 

\begin{figure}
\begin{center}
\includegraphics*[width=0.65\textwidth]{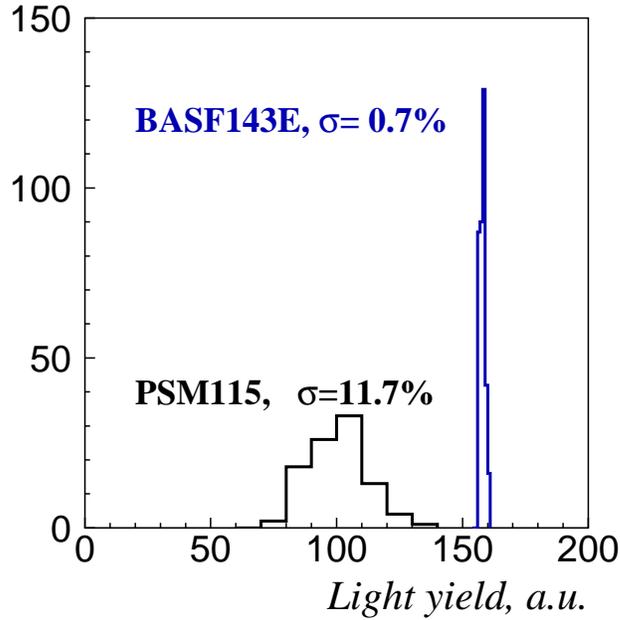}
\caption {Comparison of the light-yield spread for scintillator 
         tiles used in earlier (PSM115) and new (BASF143E) Shashlyk modules.}
\label{tile_light_spread}
\end{center}
\end{figure}

The new plastic base of the scintillator material and the new
production technology of tiles yields a 97\%
reflection efficiency from the scintillator surface, and 
strongly improves the reproduction quality (see
Fig.~\ref{tile_light_spread}) of the tiles. The latter is crucial for 
good longitudinal light collection uniformity.  Due to the fluctuation of
the depth of the electromagnetic shower in the calorimeter module, good
reproduction quality of the tiles is a mandatory condition for an
appropriate performance of the $3\%/\sqrt{E\ \mathrm{(GeV)}}$ module.

\begin{figure}
\begin{center}
\includegraphics*[width=0.65\textwidth]{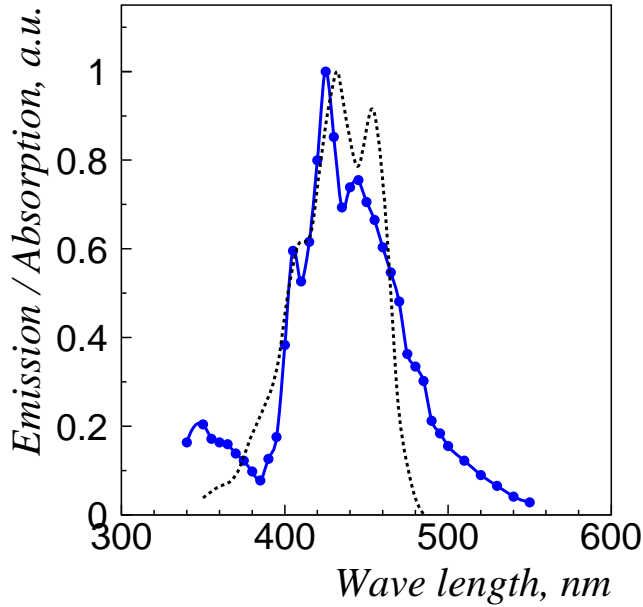}
\caption{The closed circles and solid line are the measured emission 
         spectra of the BASF143E-based scintillator. The dashed line 
         is the absorption spectrum of KURARAY Y11(200)MS WLS fibers.}
\label{abs_emit_2}
\end{center}        
\end{figure}

Fluors for the new scintillator composition, BASF143E polystyrene +
1.5\% PTP + 0.04\% POPOP were selected~\cite{scint1,scint2} 
to match well with the absorption spectrum of the Kuraray Y11(200)MS 
WLS fiber (Fig.\ \ref{abs_emit_2}).
With a BASF143E-based scintillator and KURARAY fibers, 
the effective light yield in the KOPIO Shashlyk 
module (at the entrance to the photo-detector) becomes $N_{\gamma
}\approx\ 60$ 
photons per 1~MeV of the incident photon energy. 

The radiation hardness of the scintillator may be a limiting factor
for using the Shashlyk calorimeter in the modern high statistics experiment.
It was reported \cite{scint1} that the radiation stability of 
BASF143E-based scintillator is a dose level 120 krad, and the recovery 
time is about 80 days. This limitation would not be a problem for the KOPIO
Calorimeter for which this Shashlyk module was designed.

\subsection{Photo-detector}

The Avalanche Photo Diode, a detector with high quantum efficiency, 
provides another important improvement of the photo-statistics contribution 
to the energy resolution of the Shashlyk module.

Our previous experimental studies of fine-sampling Shashlyk
modules~\cite{NIM_A531_467} have shown that the performance of
Shashlyk calorimeter modules with photomultiplier (PMT) readouts 
agrees with the simulations and nearly satisfies the requirements
formulated in the introduction.
However, these test measurements were made in 
an optimal environment for PMT readout: no magnetic fields, short 
measuring runs, and continuous and stable electron beams.

These conditions do not properly represent those that would be encountered 
in the KOPIO experimental environment. 
Our main concern is related to the leakage 
of the magnetic field from the downstream magnet of up to 500 Gauss.
Another serious problem is the cycling of real beam every 
several seconds, with possible significant variations in intensity.
This will produce short- and long-term variations of the PMT 
gain, which will destroy the response stability of the detector performance.
In addition, the quantum efficiency of PMTs, even for the best
``standard'' green-sensitive tubes, e.g. PMT 9903B of Electron Tube
Inc., is relatively low, about 20\%  at 500 nm
(the region of WLS-fiber response).

\begin{table}
\caption{Some experimental and catalog parameters of
large-area APDs.  
Quantum efficiency is given for green light, $\lambda=500\ \mathrm{nm}$.
Signal rise and fall time are measured without amplifier. 
Excess noise factor, which approximately depends on
the gain $M$ as $2+kM$ is given for $M=100$.
$I_S$ and $I_B$ are components of the dark current $I_d=I_S+I_BM$.
The gain dependences on temperature, $T$, and high voltage, $V$, are
given for $M\sim100$.
For RMD APD, we have observed a significantly different results for values of series
resistance (and consequently for signal fall time) which is indicated
in the Table. 
}
\label{Tab1}
\begin{tabular}{lcccc}
\hline
Manufacturer                & API           & RMD          & EG\&G        & Hamamatsu    \\
Model number                & 630-70-74-510 & S1315        & C30703E      & S8664-1010N \\
Active area, $\mathrm{mm}^2$ & 200           & 169          & 100          & 100          \\
                            & $\oslash16$   & $13\times13$ & $10\times10$ & $10\times10$ \\
Quantum Eff., \%            & $      94 $   & $      65 $  & $      75 $  & $      80 $  \\
Capacitance, pF             & $     130 $   & $     110 $  & $      80 $  & $     270 $  \\
Series resistance, $\Omega$ & $      15 $   & $    50 (    280)$ 
                                                           & $      10 $  & $< 5      $  \\  
Signal rise time, ns        & $       6 $   & $       9 $  & $       5 $  & $      4.4$  \\
Signal fall time, ns        & $      16 $   & $    24 (     80)$
                                                           & $      15 $  & $       25$  \\
Gain                        & $\leq 600 $   & $\leq 1000$  & $\leq 200 $  & $\leq 300 $  \\
Excess noise factor         & $     2.2 $   & $     2.05$  & $     3.2 $  & $     2.5 $  \\
$I_S$, nA                   & $      50 $   & $      310$  & $      90 $  & $       3 $  \\
$I_B$, nA                   & $     0.6 $   & $      3.8$  & $     0.4 $  & $     0.3 $  \\
$\frac{dM}{M\,dT}$, ${}^\circ\mathrm{C}^{-1}$
                            & $    -0.03$   & $    -0.03$  & $    -0.04$  & $    -0.045$ \\
$\frac{dM}{M\,dV}$, $\mathrm{V}^{-1}$
                            & $     0.014$  & $     0.016$ & $     0.03$  & $     0.03$  \\

\hline
\end{tabular}
\end{table}

Consideration was thus given to alternative photo-detectors. Recent
progress in the development of new  
APDs with large active areas (up to 
200~mm$^2$), low capacitance, low dark current, high gain (up to 1,000), 
and high quantum efficiency (up to 90\%) allows us to consider 
these photo-detectors as primary candidates for the improved Shashlyk
Calorimeter. We have studied the large area ($\geq100\ \mathrm{mm}^2$)
APDs of {\em (i) Advanced Photonix Inc.} (API), {\em (ii) RMD
  Instruments Inc.} (RMD), {\em (iii) Perkin-Elmer Inc.} (EG\&G), and
{\em (iv)   Hamamatsu Photonix K.K}.  
Some characteristics of these APDs are summarized in Table~\ref{Tab1}. 
Here, we briefly report the results of our study.

\begin{itemize}
\item Size of the active area. \\
The fibers from the Shashlyk
module, described in this paper, are collected in a 14 mm diameter
bunch. This size is well 
matched to the API APD, 
which has a diameter of 16 mm. Optical light guides must be be used with
other APDs that were considered, to match the size of the active area. For
$10\times10\ \mathrm{mm}^2$ active area, this will result in a light
loss of at least 10-15\%.

\item Response uniformity of the active area.\\
 This feature is important
for the Shashlyk module because each fiber delivers light to only a
small part of the total sensitive area.  The response of the selected
APD varies by less than 3\% over the active area. Usually, PMTs
exhibit poor spatial response uniformity ($\ge20\%$).

\begin{figure}
\begin{center}
\includegraphics*[width=0.65\textwidth]{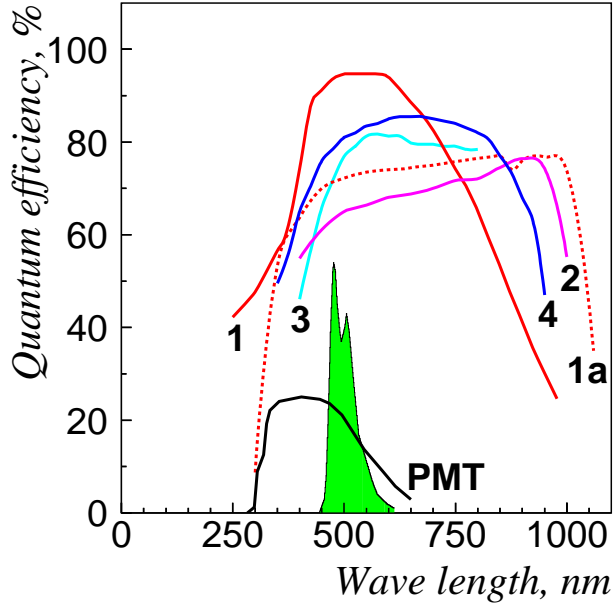}
\caption {
  The quantum efficiency of considered APDs: 
  $1-$  API ($\mathrm{SiO}_2$ window), 
  $1a-$ API (epoxy window), 
  $2-$  RMD (epoxy window),
  $3-$  EG\&G ($\mathrm{SIN}_x$ window),
  $4-$  Hamamatsu (epoxy window).
  For comparison, quantum efficiency of the 9903B PMT is also shown.
  The histogram is an emission spectrum of the Kuraray Y11200MS
  WLS fiber.
} 
\label{apd_qe}
\end{center}
\end{figure}

\begin{figure}
\begin{center}
\includegraphics*[width=0.65\textwidth]{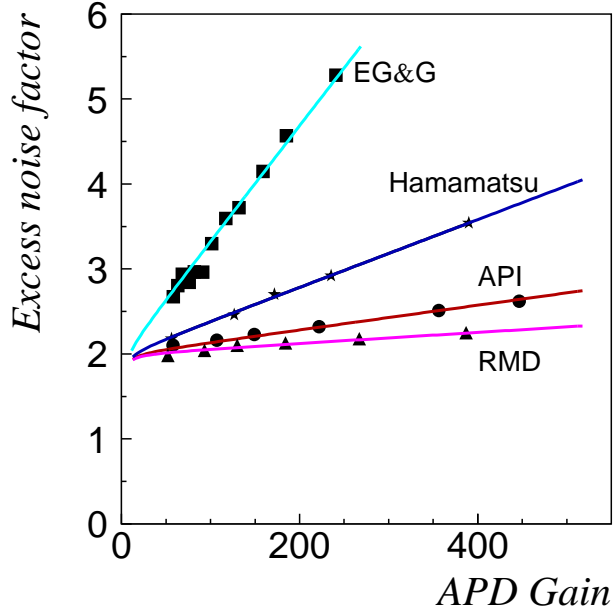}
\caption{The APD excess noise factor {\em vs.} the APD gain.}
\label{apd_f-factor}
\end{center}
\end{figure}

\item Quantum efficiency of the sensitive area.\\
Photo-statistics, directly dependent on the quantum efficiency ($Q$) of the photo-detector, 
is an important contribution to the energy resolution of a
calorimeter. For green light emitted by Kuraray fibers, the 
APDs have much higher quantum efficiencies than PMTs (see
Fig.~\ref{apd_qe}). For example, the quantum efficiency of the API APD is
94\%, a factor 5 higher than quantum efficiency of the 9903B PMT. It
should be noted that that the photo-statistics contribution depends
also on the fluctuations of the photo-detector gain, 
$(\sigma_E/E)^2_\mathrm{ph.stat.}= F/QN_\gamma$, where F is the so-called
excess noise factor and $N_\gamma$ is the number of primary photons at
the entrance of the photo-detector. For an ideal photo-detector, $F=1$,
for a high-linearity PMT it is usually between 1.3 and 1.6. 
For an APD, excess noise factor dependences on gain $M$ may be well
approximated as $F \simeq 2 + kM$ if $M\geq20$ \cite{APD-th}. The experimentally measured 
behavior of $F$ versus $M$ for the
APDs under consideration is shown in Fig.~\ref{apd_f-factor}. Taking into
account both quantum efficiency $Q$ and excess noise factor $F$, we
conclude that API APD has best ``photo-statistics quality'' (
$Q/F\simeq43\%\ (M_\mathrm{APD}\simeq100)$) of about a factor 3.5 better 
than 9903B PMT.

\item Electronic noise of the photo-detector/amplifier chain.\\ 
In the case of a Charge Sensitive Amplifier (CSA) with a simple 
$CR$-$RC$ shaper, the contribution of the electronic noise from the photo-detector/amplifier chain to the energy resolution can be estimated as: 

\begin{equation} 
\sigma_\mathrm{noise} = \frac{1}{n_\gamma Q} \sqrt{\frac{I_S/M^2 + I_BF}{q}\Delta t  +  \frac{\sigma^2_{\mathrm{amp}}(C,\Delta t)}{M^2}}
\label{eq:2}
\end{equation}
where $n_\gamma$ is number of photons at the entrance of photo-detector
per unit of the deposited energy ($n_\gamma\approx60\
\mathrm{MeV}^{-1}$ for the improved Shashlyk module),
$q=1.6\cdot10^{-19}\ \mathrm{C}$ is electron charge,
and $\Delta t$ is the measurement gate width. The photo-detector
noise can be related to the $I_S$ (surface leakage current) and $I_B$ (bulk
current), the components of the dark current $I_d=I_S+I_BM$. The
amplifier noise, $\sigma_\mathrm{amp}$, defined here as Equivalent
Noise Charge (ENC) depends on the gate width $\Delta t$ and the
photo-detector capacitance $C$.

\begin{figure}
\begin{center}
\includegraphics*[width=0.65\textwidth]{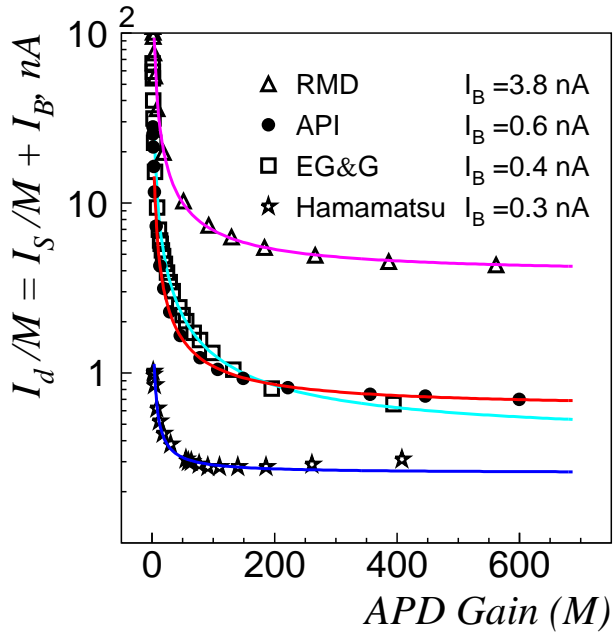}
\caption{The APD dark current.}
\label{apd_dc}
\end{center}
\end{figure}

For a PMT readout, the photo-detector noise is
negligible due to a small ($\sim0.1\ \mathrm{nA}$) PMT dark current
and big ($\geq10^5$) gain. The
experimentally measured behavior  
of the APD dark current versus gain $M$ for our 
APDs is shown in Fig.~\ref{apd_dc}. The contribution of $I_S$ to the
noise becomes negligible if $M\geq100$.  For the API APD, the
photo-detector contribution to the noise can be estimated as 
$0.5\ \mathrm{MeV}$  
for $\Delta t\approx 100\ \mathrm{ns}$, $M \geq 100$, and
environmental temperature $T=25\ {}^\circ\mathrm{C}$.
  
The value of the amplifier noise $\sigma_{\mathrm{amp}}$ is usually 
measured experimentally. We have used a fast low noise CSA, that has
been designed for KOPIO Calorimeter to optimize the signal-to-noise
ratio and double-pulse resolution. Its charge-sensitive part was
designed as a cascade amplifier with two parallel-connected low
capacitance JFET transistors.  We considered the Russian KP341A and
Japanese SK2394/YJ5 transistors for this amplifier. Both transistors
are characterized by a high transconductance ($g_m>20\ \mathrm{mA/V}$)
for moderate noise current and low input capacitance
($C_\mathrm{JFET}<10\ \mathrm{pF}$). For a $100\ \mathrm{ns}$ gate,
the ENC dependence on the 
photo-detector capacity $C\ \mathrm{(pF)}$ was measured to be $270+18.8C$
for the KP341A and $500+10.9C$ for the SK2394/YJ5. We used an amplifier with
two SK2394/YJ5 transistors in the test beam measurements. For the API APD
($C=130\ \mathrm{pF}$) the amplifier noise may be estimated as
$0.3\ \mathrm{MeV}$ resulting in a total photo-detector/amplifier
chain noise of $0.6\ \mathrm{MeV}$ for $\Delta t=100\
\mathrm{ns}$, $M=100$, and $T=25\ {}^{\circ}C$.
It should be underlined, that large photo-statistics, $n_\gamma$,
is a crucial factor in achieving required effective noise suppression.

\begin{figure}
\begin{center}
\includegraphics*[width=0.65\textwidth]{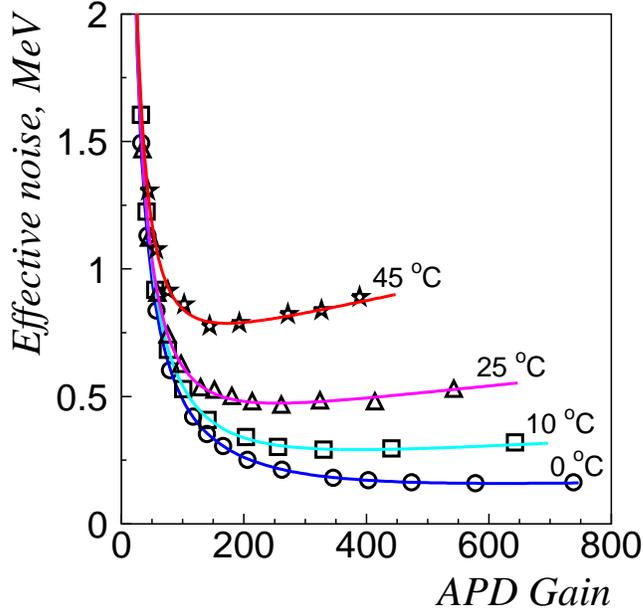}
\caption{
  Effective noise of the API APD/amplifier chain {\em vs.} the 
  APD gain for various environmental temperatures.
} 
\label{apd_noise}
\end{center}
\end{figure}

\item Gain and noise dependence on the temperature.\\
The APD performance is very sensitive to the environmental
temperature. Because $\frac{1}{M}\frac{dM}{dT}\simeq -4\%$, 
temperature and gain monitoring is important in 
using APDs in the experiment.  
For the API APD, the noise dependence on temperature is
shown in Fig.~\ref{apd_noise}. 
A thermostable cooling system and appropriate increase
of the APD gain may allow one to lower the effective noise down to
the level of 0.2-0.3 MeV.

\item Gain dependence on High Voltage.\\ 
According to our experimental
study, $\frac{1}{M}\frac{dM}{dV}\approx (2-3)\%$
depending on the environment temperature and APD gain. The dependence is
similar to those of PMTs.

\item Gain dependence on the rate.\\ 
Contrary to PMTs, there is almost
no dependence of APD gain on rate. In our test measurements
with API APDs, we did not find any gain variations for the rates up to
few $\mathrm{MHz}$. However, APDs are used with preamplifiers. For
slow preamplifiers, there may be a rate dependence due to the pile-up. 

\begin{figure}
\begin{center}
\includegraphics*[width=0.5\textwidth]{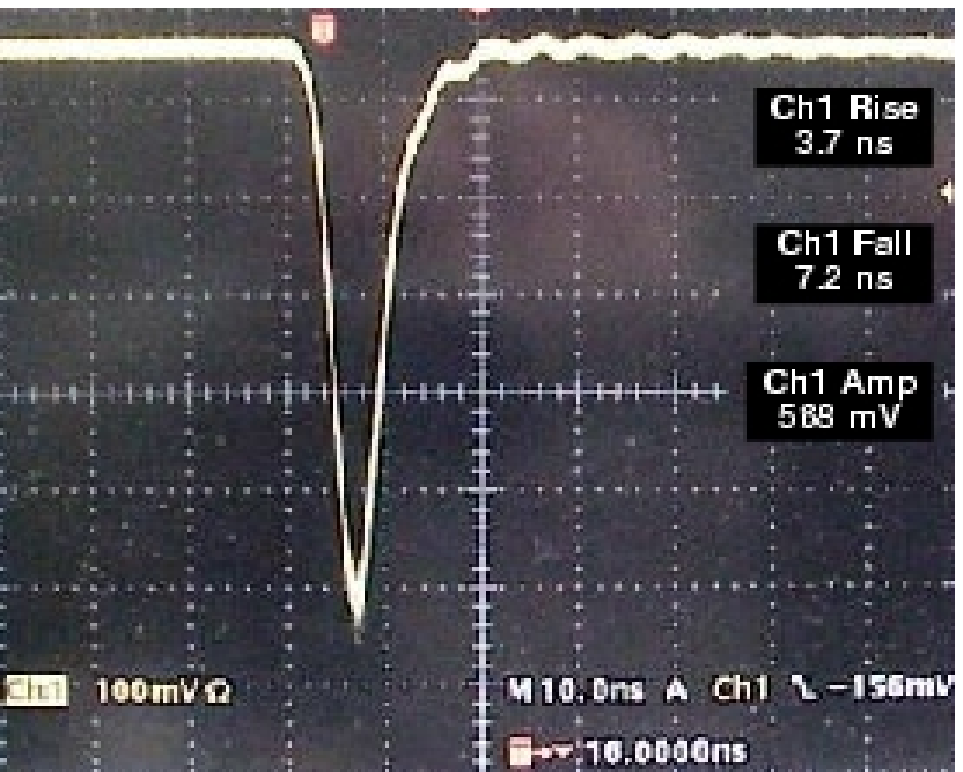}
\includegraphics*[width=0.5\textwidth]{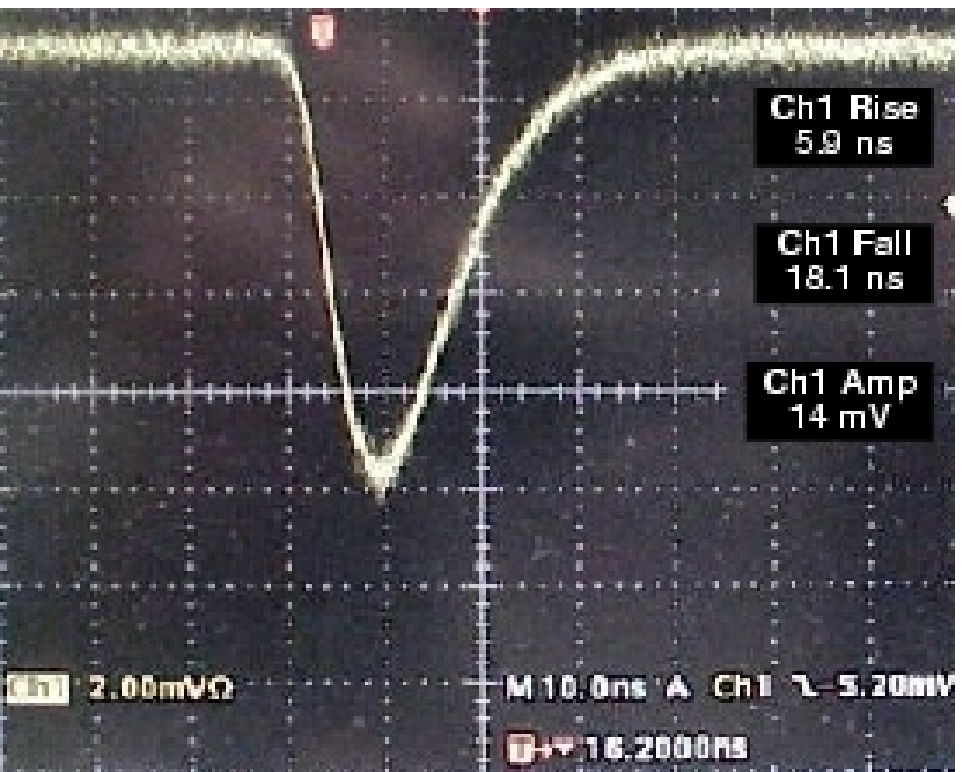}
\caption {
  PMT-9903B (upper) and API APD (lower) responses to 
  short, 3 ns, light pulses. 
  The intensity of the light pulse corresponds to the 700 MeV photon signal.}
\label{pmt-apd_response}
\end{center}
\end{figure}

\item Time response of the API APD.\\ 
The time response of the API APD
(without amplifier) to a short-duration (3 ns) light 
pulse with an intensity corresponding to a 700-MeV photon, is shown 
in Fig.~\ref {pmt-apd_response}.  The APD rise time of 5.9 ns is
comparable with 3.7 ns for the 9903B PMT  
response to the same light pulse.  Due to higher APD capacitance, the fall 
time of the APD pulse, 18.1 ns, is longer than that for the PMT response of 7.2 ns.
  
The difference between a PMT and an APD time response becomes less important 
if we take into account the Shashlyk module signal (light pulse) duration. 
(see Fig.~\ref{wfd_signal}). However, both
the photo-detector capacitance $C$ and serial resistance $R_s$ are
important parameters for the time response of the APD.

\item Operation in magnetic field.\\
As opposed to PMTs,  APDs can operate in a magnetic field of up to 80 kG~\cite{APD_magnet}.

\end{itemize}

From the above consideration of the available APDs, we conclude that
the Advanced Photonix Inc.\ APD 630-70-74-510 is the optimal choice for
our goals.
Having the largest active area which matches well with the
fiber bunch size, this commercial APD has best photo-statistics
quality, $Q/F$, and sufficiently short time response when compared to other large area APDS . The relatively
high  gain and low dependence of
 excess noise factor on gain, allows us to optimize detector
performance, e.g.,  to reduce the effective photo-detector/amplifier
noise to the 0.2-0.3 MeV level for reasonable values of the
environmental temperature.

These characteristics of the API APD
together with the improved light yield in the new Shashlyk module
 reduces the photo-statistics contribution to the energy
resolution of the calorimeter to a negligible level of
$0.7\%/\sqrt{E\ \mathrm{(GeV)}}$.  The electronic noise
contribution is also negligible.

\section{Experimental Study of the Calorimeter Prototypes}

\subsection{Test beam}

Test measurements of the prototype of a Shashlyk calorimeter with 
energy resolution of $3\%/\sqrt{E\ \mathrm{(GeV)}}$ have been made 
with the photon beam from the Laser Electron Gamma Source (LEGS) 
facility ~\cite{LEGS}.  LEGS is located at the National
Synchrotron Light Source (NSLS) at Brookhaven  
National Laboratory.
A continuous photon beam with an energy range of 150 to 400 MeV
is produced by laser photons Compton backscattered by 
2.58-GeV storage ring electrons.
The energy of each backscattered photon was known by detecting the Compton
scattered electron. 

The tagged photon beam had an average intensity of $\sim3\times10^5$ Hz,
 horizontal and vertical sizes of $\sim$1.5 cm at the detector position, and an angular 
spread of $\sim$2 mrad.  The photon energy was tagged with an 
accuracy of $\delta E_\gamma/E_\gamma\approx1.5\%$.  
The timing accuracy of the tagging spectrometer was $\simeq$1 ns.  
Several monochromatic photon energy lines in the range of 220 MeV to 370 MeV 
were triggered in our measurements.

\subsection{Calorimeter Prototypes}

\begin{figure}
\begin{center}
\mbox{\includegraphics*[angle=-90,width=0.8\textwidth]{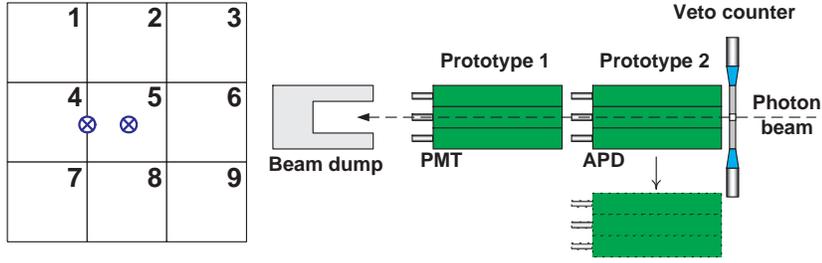}}
\end{center}
\caption{Schematic drawing of the detector 
         setup. An indexing of the nonet modules and a beam position at 
         a studied nonet for the energy, inefficiency (center), and
         timing (between modules 4 and 5) measurements is shown on the
         left.}
\label{Beam-setup}
\end{figure}

Two arrays of $3\times 3$ prototype modules with the same sampling of
275 $\mu\mathrm{m}$ of lead and 1.5 mm of scintillator  
were tested as shown in Fig. \ref{Beam-setup}.

The first array (prototype 1) contained 9 module with paper between
lead and scintillator (earlier design) with 
the 30-mm-diameter, green-sensitive PMT-9903B of {\em Electron Tube 
Inc.}  

\begin{figure}
\begin{center}
\includegraphics*[width=0.65\textwidth]{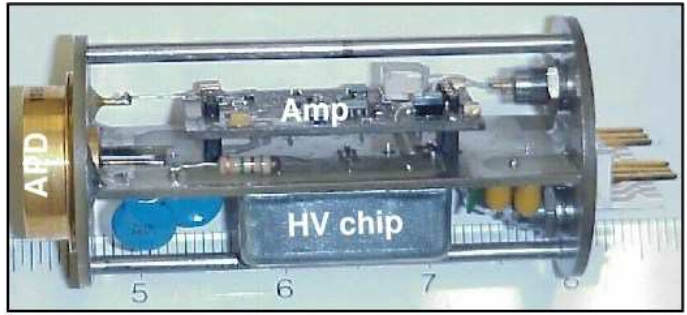}
\caption{View of an APD unit.} 
\label{apd_unit}
\end{center}
\end{figure}

The second array (prototype 2) contained a nonet of new design modules 
equipped with the 16 mm diameter API APD. 
 The APD detector housing, an instrumented unit 
including the APD itself, the APD amplifier, and 
the APD HV bias, is shown in Fig.~\ref{apd_unit}.

Because the new Calorimeter photo-detector (APD) draws a typical photo-current
of less than 0.1 mA, an economical way to build the 
Calorimeter HV system is to develop an active, compact  
individual HV-unit mounted directly into each APD housing. This kind 
of HV system eliminates expensive and bulky HV cables and connectors,
lowers the power consumption per individual power supply, and reduces 
the electrical HV hazard associated with traditional HV supplies.

In the test measurements, two types of APD HV supplies were tested: 
traditional HV supply: a NIM standard {\em BERTAN} 377P,  and a HV  ``built-in" unit 
(see Fig.~\ref{apd_unit}) with a new commercial, compact, regulated and programmable
LV-HV converter (C20), produced by EMCO. The stability of both 
bias systems was better than 0.1 V/hour, which provided an APD gain 
stability better than 0.3\%.

\subsection{Cooling}

The photo-detectors (APD) of the Prototype 2 were placed in a thermo-isolated cooling unit.
During the test period of 20 hours, the APD temperature was kept at
$18\ ^\circ\mathrm{C}$.  Variation did not exceed $0.2\ ^\circ\mathrm{C}$. Such cooling
system allowed us to neglect the temperature dependent effects in data analysis. 

\subsection{Readout}

One of the goal of the photon beam measurements was to test the
readout prototype for the KOPIO experiment.
The readout electronics had to be capable of
measuring energies with a digitization uncertainty of 0.5 MeV in a
dynamic range of 2--1000 MeV, and the time of arrival with respect to
the beam micro-bunch clock with an uncertainty of less than a few hundred ps. 
Operation of this readout electronics was required to be completely pipelined
with no dead time.  
It also had to be capable of forming and discriminating the total
energy signals and the total number of hits in the Calorimeter. 
To meet these performance requirements for energy and timing
resolution with PMTs or APDs, in an economical way,
we have investigated using Wave Form Digitizers (WFDs).  Our WFD was based on an8-bit 140-MHz WFD which
has been developed by Yale University \cite{WFD}. 
Its measurements were compared with
  those of conventional 12-bit LeCroy QDCs.

\subsection{Calibration and Monitoring}

The energy resolution of 
$\sim$3.0\%/$\sqrt{E}$  requires that a pre-calibration and monitoring system 
be developed.

We have used vertical cosmic-ray muons for pre-calibration of
the calorimeter prototypes, as we did in 
Experiment E865~\cite{NIM_A479_349}. The accuracy of this pre-calibration
was estimated to be about 4\%. 

To monitor prototype performance during our
measurements and to correct the variation of the photo-detector
gains with an accuracy of 0.4\%, we have used a prototype of the KOPIO
Calorimeter monitoring system.  This system
employs high-brightness Light Emitting Diodes (LEDs NSPB-310A)
to inject blue light into the scintillator via a clear fiber
(see Fig. \ref{module_design}).
Stabilization
of the LED light was obtained by means of an optical feedback provided by a 
PIN photo-diode as shown in References \cite{PIN1} and~\cite{PIN2}.

This system satisfied a number of our specific requirements:
\begin{itemize}
\item short light pulse duration, less than 15 ns;
\item variable light pulse intensity ($5,000\mathrm{-}20,000$ photons/channel);
\item variable pulse repetition rate (up to $1MHz$);
\item high long-term and short-term temperature stability, 
better than 0.1\%;
\item small variation in the flash amplitude, less than 0.2\%.
\end{itemize}
The monitoring system can also serve as 
a pre-calibration of the module readout chain at the 10\% level.

For final calibration of the prototypes we used
the photon beam.
Both calorimeter prototypes were mounted on platforms that could 
be moved horizontally and vertically with respect to the beam line,
so that each prototype module could be calibrated (with an accuracy 
$\leq 1\%$) by using the 250-MeV photon beam that passed through 
the central region of the module at normal incidence. 

\section{Experimental Results}

\subsection{Calorimeter prototype response}

The pedestal response, the sum of electronic noise and pile-up for 
individual modules, was measured during the test runs by using a 
special gate signal that was shifted from the photon timing by up 
to 1 $\mu$s. The contribution of this effective noise to the energy
resolution was $(0.5\pm 0.1)$ MeV for the case of the ``APD+QDC'' 
readout. The total equivalent noise for the ``APD+WFD'' readout was 
$(1.0\pm 0.2)$ MeV. This latter value is twice as large as that for the 
first one due to a digitization uncertainty of the 8-bit WFD. 
The lowest equivalent noise, $(0.2\pm 0.1)$ MeV, was obtained with 
the PMT-9903B tubes.

The typical response of the nonet of the ``APD+WFD"-instrumented
modules to a 340 MeV photon hitting the nonet at the center of the
central module, is shown in Fig.~\ref{wfd_signal}. The measured
signal distribution in the WFD was fit by a function $Af(t-t_0)+P$, where
$f(\tau)$ is an experimentally determined pulse shape function. The
signal amplitude $A$, signal time $t_0$, and pedestal $P$ were free
parameters in the fit.

\subsection{Energy Resolution}

\begin{figure}
\begin{center}
\includegraphics*[width=0.65\textwidth]{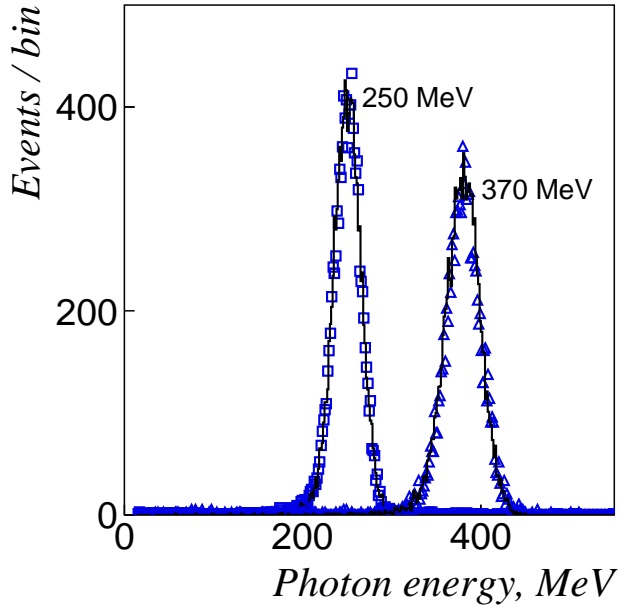}
\caption{
  Comparison of the experimental (marks) and 
  simulated (solid lines) energy spectra in the Shashlyk prototype
  with APD+WFD readout.
}
\label{shashlyk_spectra}
\end{center}
\end{figure}

The energy resolution was measured for both prototypes 1 and 2. The
photon beam was directed to the center of the module nonets.
Energy spectra of 250- and 370-MeV photons in the Shashlyk nonet 
with the APD+WFD readout are shown in Fig.~\ref{shashlyk_spectra}. 
Note the good agreement between Monte-Carlo and experimental 
distributions, that gives us a confidence that our simulation
model, which includes effects of the beam energy spread,
energy loss upstream calorimeter,
photon shower evolution, light collection in scintillator tiles and light 
transmission in WLS fibers, the response of the photo-detector, and 
noise of the entire electronic chain, properly reproduces the actual calorimeter response.

\begin{figure}
\begin{center}
\includegraphics*[width=0.65\textwidth]{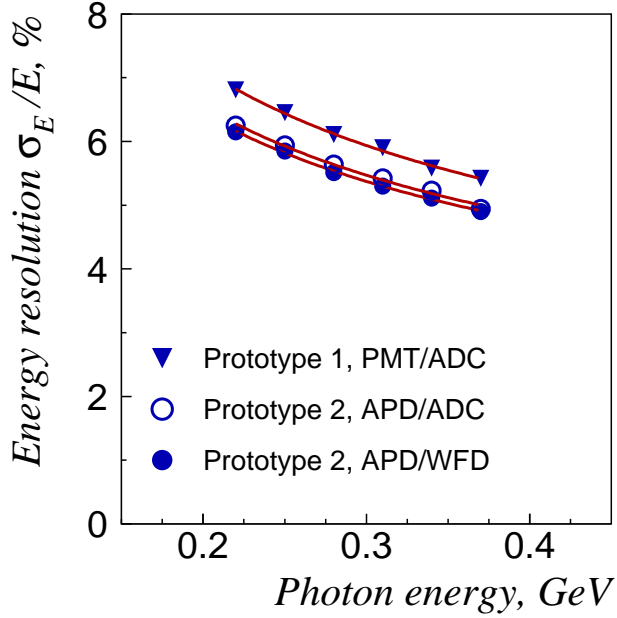}
\caption{Energy resolution of a prototype of the Shashlyk Calorimeter.} 
\label{e_res}
\end{center}
\end{figure}

The energy-resolution results for various readouts
are displayed in Fig.~\ref{e_res}. 
The best result was achieved for the ``APD+WFD" readout.  A fit to 
these experimental data gives
\begin{equation}
\sigma_E/E=(1.96\pm0.1)\%\oplus(2.74\pm0.05)\%/\sqrt{E\ \mathrm{(GeV)}}\:,
\end{equation}
where $\oplus$ means a quadratic summation. The relatively large 
constant term of 2\% may be explained by the short, 15.9-X$_0$ 
radiation length of the module. The constant term may be
improved by increasing length of the module. However, this term is not
essential for the photon energy range of $50\mathrm{-}1000$ MeV.

\subsection{Time Resolution}

To estimate the time resolution, we measured the time difference for 
the signals produced by the same shower in two neighbor modules.
By using this measurement technique, the 340 MeV photon beam was 
directed between central module (\#5) and left module (\#4), 
see Fig. \ref{Beam-setup}.

\begin{figure}
\begin{center}
\includegraphics*[width=0.65\textwidth]{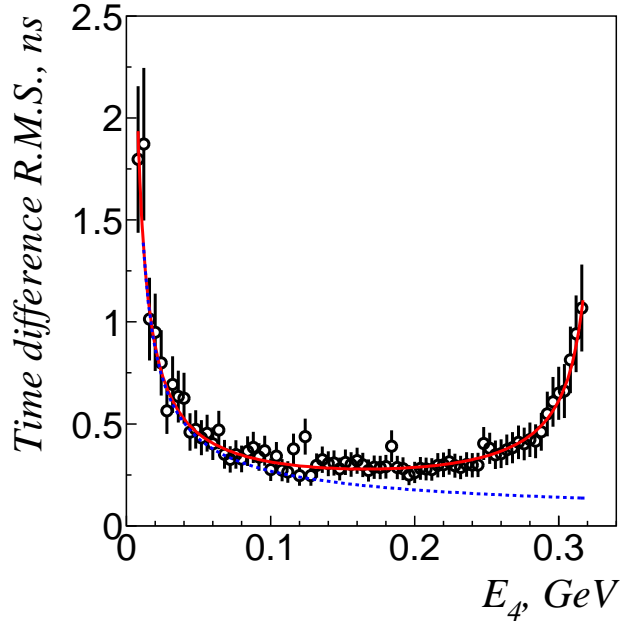}
\caption{
  Experimental evaluation of the time resolution of a Shashlyk
  module (see text for details). $E_4$ is energy measured in module \#4.
  Solid line is an expectation of the time difference R.M.S.
  $\sigma_{(T_4-T_5)}(E_4)=\sqrt{\sigma^2_T(E_4)+\sigma^2_{T}(0.33-E_4)}$,
  where 
  $\sigma_T(E)=0.072/\sqrt{E}\oplus0.014/E$ (shown by dashed line)
  is a fit-estimated time resolution in each module, \#4 and \#5.
}
\label{t_res-2}
\end{center}
\end{figure}

Only events with the full photon energy ($E_4 + E_5$ = $330\pm 20$ MeV) 
deposited in these two modules were selected for analysis.
The dependence of the time difference variance on the energy deposited 
in  module \#4 is shown in Fig.~\ref{t_res-2}. The time difference 
resolution drops significantly if one of the two deposited energies is 
close to zero or, alternatively, if one of the energies is close to 
330 MeV and the other is necessarily close to zero. 
Assuming that both modules have the same time resolution, 
we have obtained from Fig.~\ref{t_res-2}
\begin{equation}
\sigma_T = \frac{ (72\pm4)\ \mathrm{ps} }{ \sqrt{E\ \mathrm{(GeV)}}}
           \oplus
           \frac{ (14\pm2)\ \mathrm{ps} }{ E\ \mathrm{(GeV)}}           
\end{equation}

The signals we use to evaluate the time resolution are strongly
correlated because they are produced by the same electromagnetic
shower. This may result in a wrong value of the time resolution. 
For example, the contribution of the longitudinal fluctuation
of the shower is suppressed in such measurements. To ensure that our
test-beam result does not underestimate the actual time resolution, we
carried out a Monte Carlo simulation of the time resolution.

To simulate the time response and time resolution of a Shashlyk
module, a Monte Carlo model was upgraded for the evolution of the 
light signal in a module. Ionization produced by a charged particle 
in a scintillator tile survives several transformations before the 
corresponding light signal appears at the entrance of a photo-detector.
Ionization produced by an electromagnetic shower occur at different 
space points and different times, resulting in a time spread of the 
photo-electron emission in the photo-detector. In addition, 
the actual emission delays depend randomly on the decay times in 
the scintillator and the WLS fibers, light collection time in the 
scintillator fiber, and the propagation of light in the WLS fiber.
The following effects have been taken into account in the model:
\begin{itemize}
\item the space-time distribution of hits in the Shashlyk module scintillator;
\item the decay time in scintillator;
\item the light collection time in the scintillator tile; 
\item the decay time in WLS fibers; 
\item the effective velocity of light propagation in WLS fibers;
\item the light attenuation length in WLS fibers; 
\item bending loss in the WLS fiber loop; and 
\item the response function of the detector chain, including
photo-detector (APD), preamplifier, cables and WFD.
\end{itemize}

\begin{figure}
\begin{center}
\includegraphics*[width=0.65\textwidth]{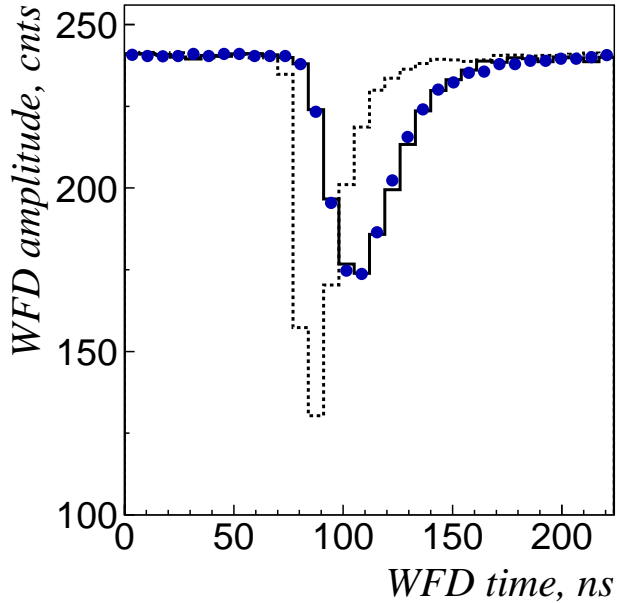}
\caption{
  A 340 MeV photon signal in WFD. Closed circles are experimental data. 
  Solid line is the simulation, including readout chain response function. 
  Dashed line is the light pulse shape at the entrance of the APD. 
}
\label{wfd_signal}
\end{center}
\end{figure}

The simulated response of the Shashlyk module for the 340 MeV photons 
is shown in Fig.~\ref{wfd_signal}. The shape of the simulated
signal in a Wave Form Digitizer matches well an experimentally 
measured signal.

In the Monte-Carlo simulation of our time resolution  measurements, 
we have obtained $\sigma_{T4-T5}= 280\ \mathrm{ps}$ if $E_4\approx E_5$
which is in a good agreement with the experimental value of 285 ps 
(see Fig.~\ref{t_res-2}).
 
\begin{figure}
\begin{center}
\includegraphics*[width=0.8\textwidth]{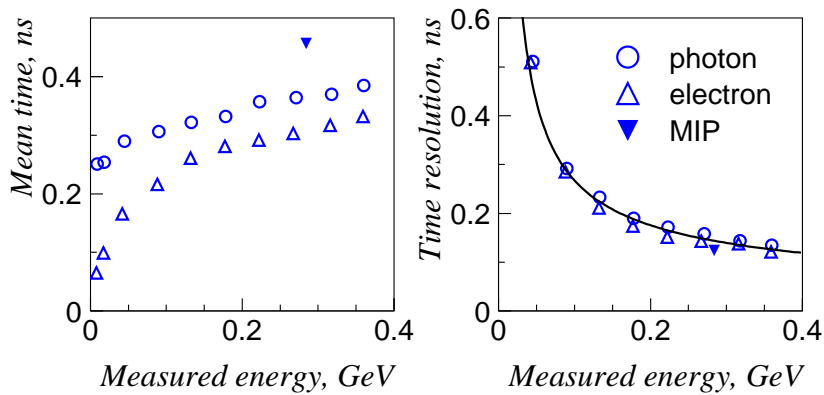}
\caption{
  Results of a Monte Carlo simulations of the time resolution of 
  the Shashlyk module for photons, electrons, and Minimum Ionizing
  Particles (MIP). 
  The solid line corresponds to a 
  $0.072/\sqrt{E}\oplus0.014/E\ \mathrm{(ns)}$, 
  an approximation of the time resolution from our test-beam measurements.
  The simulated dependence of the mean measured time on type
  of particle and its energy is shown on the left. 
}
\label{time_mc}
\end{center}
\end{figure}

The Monte-Carlo simulations of the mean measured time and 
time resolution  dependence on type of particle and its energy is shown
in Fig. \ref{time_mc}. To understand the behavior of mean time
values we should note that by fitting the signal shape in a WFD we
actually measure the center of gravity of the signal time
distribution.

The main contributions to the relative signal delay are electron/photon
tracking time before the hit and  time of light propagation in the
fiber.
 The average of time delays in both ends of the fiber does not
depend on the 
actual position of the hit if we can neglect the light
attenuation in the fiber. In such an approximation, the
relative delay of the measured hit is solely defined by the time at
which this hit occurred. In other words, the deeper shower penetrates in a
module the larger time delay is measured.
Since photons interact more deeply in the module than electrons, 
the measured time
for photons is expected to be about 100 ps larger that for electrons of the
same energy.

It is interesting to note that one can 
suppress the dependence of the measured time on the type of particle and
its energy if the loss of light 
in the loop were equal to
$2v/(c+v)\approx0.7$, where $c$ is the speed of light in vacuum and $v$ is the
effective velocity of signal propagation in fiber. 
However, such a degradation of
the loop would reduce the photo-statistics by about 30\%, and also
the contribution to the shower longitudinal fluctuations will be increased.

According to the Monte-Carlo simulation, the Shashlyk module time
resolution obtained by our method, based on the measurement of time
difference in 
the neighbor modules, is consistent with 
the actual time resolution of photon/electron detection.

\subsection{Photon Detection Inefficiency}

\begin{figure}
\begin{center}
\includegraphics*[width=0.65\textwidth]{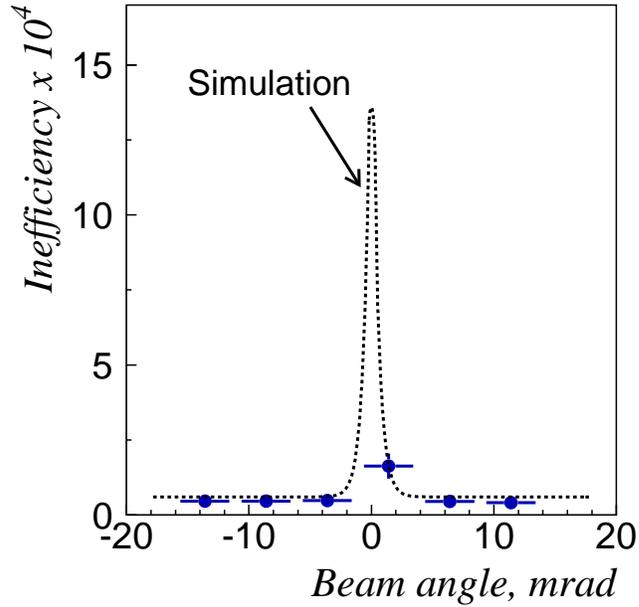}
\caption{The dependence of the photon detection inefficiency on the 
         incident angle.} 
\label{ineff}
\end{center}
\end{figure}

To estimate the photon detection inefficiency due to fiber holes, 
the prototype 1 was located behind the prototype 2 array. The absence 
of a signal in the front array, while the total photon energy was 
deposited in the second, was interpreted as a penetration of the photon 
through the first prototype without interaction, {\it e.g.} through a fiber 
hole. Photons of 250 MeV were directed onto the face of the first array.
The measured dependence of the photon detection
inefficiency on the incident angle of the beam is compared with the
simulation in Fig.~\ref{ineff}. The relatively coarse step in 
the measured angles does not allow 
us to accurately compare the experimental results with the
simulation. However, our measurements clearly indicate that the effect 
of the WLS fiber holes is negligible if the incident angle is $\ge 5$ 
mrad. For such angles, our experimental results agree well with the 
probability of photon interactions in about 8.25 cm of lead and 45 cm 
of scintillator. It has to be pointed out that the  measurement
of photon detection inefficiency in this way is insensitive to the photon
``disappearance''  in the Calorimeter, e.g. due to photo-nuclear reactions.

\subsection{Long-term APD Stability}

\begin{figure}
\begin{center}
\includegraphics*[width=0.65\textwidth]{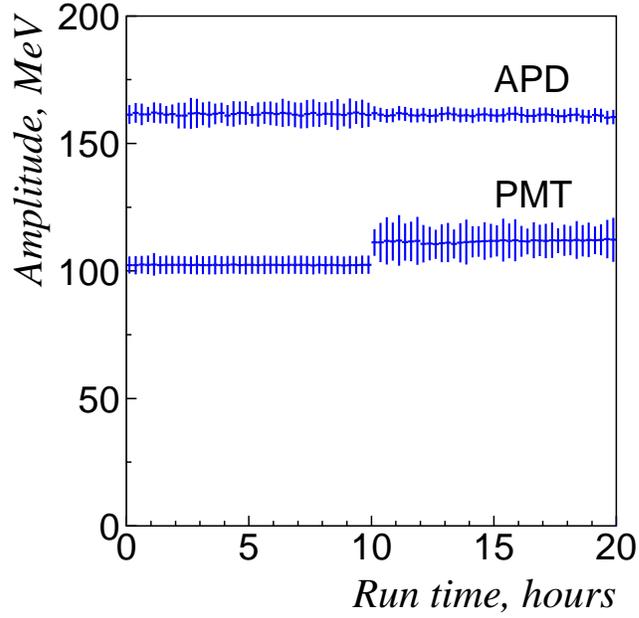}
\caption{
  Illustration of the long-term stability and gain dependence 
  on the rate for the APD and PMT readout. The mean values of the LED
  signals were measured during 20 hours run.
  The 250 MeV, 300 kHz photon beam was switched 
  from Prototype 2 with APD readout to Prototype 1 with PMT readout 
  after 10 hours of the run. Vertical lines indicate the measured
  R.M.S. of the LED signal.
}
\label{beam_intensity}
\end{center}
\end{figure}

To test the long-term stability of the photo-detector 
gain, the APD and PMT gains were monitored using LED signals over 
a 20 hour period.  Experimental data resulting from this test are shown 
in Fig.~\ref{beam_intensity}.  One can see a rate effect for the
PMT-9903B after changing the photon  
beam intensity from 0 kHz to 300 kHz. No APD gain dependence 
on the photon beam intensity was observed after changing the photon 
beam rate over this range.  
The variation of the APD gain did not exceed 1\%. 

\section{Summary}

Modules for a Shashlyk Calorimeter with energy resolution
\begin{equation}
\sigma_E/E=(1.96\pm0.1)\%\oplus(2.74\pm0.05)\%/\sqrt{E\ \mathrm{(GeV)}}\,,
\end{equation}
time resolution of
\begin{equation}
\sigma_T = (72\pm4\ \mathrm{ps})/\sqrt{E\ \mathrm{(GeV)}}
           \oplus
           (14\pm2\ \mathrm{ps})/( E\ \mathrm{(GeV)})\,,           
\end{equation}
and an inefficiency in photon detection due to the nature of the modules of
\begin{equation}
\epsilon \approx 5\times 10^{-5}\;\;(\Theta _{\mathrm{beam}} >5\;\mathrm{mrad})
\end{equation}
have been constructed and experimentally tested. The characteristics 
experimentally determined for the Calorimeter prototype well meet the 
design goals of the KOPIO experiment. 

To optimize the Calorimeter readout design, a Monte Carlo simulation 
model of the light transmission in scintillator tiles and WLS fibers,
the response of the photo-detector, and 
noise of the entire electronic chain. This model accurately describes the 
experimental data. 

\section*{Acknowledgments}
This work was supported in part by the US Department of Energy, the
National Science Foundations of the USA and Russia. We thank the
directorate of Institute for Nuclear Research (Moscow) and Institute
for High Energy Physics (Protvino) for their support of this
work. 
We are gratefully acknowledge all
participants of the KOPIO Collaboration for numerous
discussions of the Shashlyk Calorimeter performance.
We are grateful to A. Sandorfi for
the opportunity to use LEGS equipment in our test beam measurements.
We are further indebted to E.N. Guschin, Yu.V. Musienko, and P. Rehak for
many valuable discussions and suggestions.

\end{document}